\documentclass[12pt,preprint]{aastex}

\begin{document}

\title{Constraining the origin of TeV photons from gamma-ray bursts with delayed MeV-GeV
emission formed by interaction with cosmic infrared/microwave
background photons}

\author{X. Y. Wang$^{1,2}$, K. S. Cheng$^2$, Z. G. Dai$^1$  and T. Lu$^3$ }
\affil{$^1$Department of Astronomy, Nanjing University, Nanjing
210093,
China\\
$^2$Department of Physics, The University of Hong Kong, Hong Kong,
China \\
$^3$Purple Mountain Observatory, Chinese Academy of Sciences,
Nanjing 210008, China}

\begin{abstract}
It has been suggested that electromagnetic cascade of very high
energy gamma-rays from gamma-ray bursts (GRBs) in the
infrared/microwave background can produce delayed MeV-GeV photons.
This delay could be caused by the angular spreading effect of the
scattered microwave photons or deflection of the secondly pairs
due to intergalactic magnetic field. Very high energy TeV photons
of GRBs could be produced by a few mechanisms including the
proton-synchrotron radiation and electron inverse Compton emission
from GRB internal shocks as well as external shocks. We suggest
that the information provided by the delayed emission could give
constraints on models for TeV gamma-rays.  A more accurate
calculation of the delayed time caused by the angular spreading
effect is presented by considering recent observations of the
 extragalactic infrared background  and the theoretic
high-redshift infrared background. We also suggest that the
dependence of the maximum time delay of scattered photons on
their energies, if determined by future GLAST detector, could
differentiate the two mechanisms causing the time delay.

 \end{abstract}


\section{Introduction}
GeV emission from gamma-ray burst (GRB) sources is now considered
a well-established fact (e.g. Sommer et al. 1994;  Hurley et al.
1994) and there is also tentative evidence for TeV emission ({
e.g. Amenomori et al. 1996; Padilla et al. 1998}). Recently, the
Milagro group reported the detection of an excess of TeV
gamma-rays above the background from one of the 54 BATSE GRBs
(GRB970417a) in the field of view of their detector, with a chance
probability $\sim1.5\times10^{-3}$ (Atkins et al. 2000). Poirier
et al. (2002) reported suggestive evidence for sub-TeV gamma rays
arriving in coincidence with GRBs. Although these observations
were not claimed as firm detection, the production of TeV photons
are also predicted by GRB theories. The emission mechanism for TeV
photons includes electron inverse Compton (IC) emission and
synchrotron emission from the protons accelerated by GRB
shocks{\footnote{The decay of $\pi^0$'s produced in photo-meson
interactions in internal (Waxman \& Bahcall 1997) or external
shocks (Waxman \& Bahcall 2000; Dai \& Lu 2001) of GRBs could also
lead to production of very high energy photons, but with
characteristic energies higher than $10^{14}$ eV. So  a low
radiation efficiency of TeV photons is expected  for this process.
}}. The shocks could be internal shocks,  external forward shocks
or external reverse shocks of GRBs. Such very high energy photons
at cosmological distance, however, may largely be absorbed by
interacting with the cosmic infrared background radiation (CIB;
e.g. Stecker, De Jager \& Salamon 1992; Madau \& Phinney 1996).
The IC scattering of the created $e^+e^-$ pairs off the cosmic
microwave background (CMB) photons will produce delayed MeV-GeV
emission (Cheng \& Cheng 1996; Dai \& Lu 2002). There are two
likely mechanisms causing the time delay. One is the angular
spreading effect of the secondly pairs, i.e. the scattered
microwave photons deviate from the direction of the original TeV
photons by an angle $\sim 1/\gamma$, where $\gamma$ is the Lorentz
factor of the $e^+e^-$ pairs (Cheng \& Cheng 1996; Dai \& Lu
2002). Another mechanism is related to the deflection of the
propagating direction of the pairs in the intergalactic magnetic
field (IGMF), if this field is sufficiently strong (Plaga 1995).

Dai \& Lu (2002) have discussed the spectrum and duration of the
delayed emission, assuming that the high energy primary photons
(${\cal E_\gamma} >300 {\rm GeV}$){\footnote{ Please note that
here $\cal E_\gamma$ is the gamma-ray energy at the restframe of
GRBs }} are produced by the electron IC emission in internal
shocks. Recently, Guetta \& Granot (2003) argued that the
intrinsic cutoff energy of photons from internal shocks can hardly
extend to $\ga100{\rm GeV}$ for typical GRBs with peak energy in
the BATSE energy range. Here we extend the work of Dai \& Lu
(2002) by considering that TeV photons could also come from
external shocks and could be produced by proton-synchrotron
radiation. We further suggest that the information provided by the
delayed emission could constrain the emission mechanism of the TeV
photons from GRBs and distinguish between the two mechanisms
causing the time delay. We will also study the detectablity of the
delayed emission by the future GLAST detector.

In section 2, we present three emission processes for TeV photons
and the corresponding spectrum, and calculate the cutoff energy
for high energy photons from external shocks due to
$\gamma$-$\gamma$ pair attenuation with softer photons. We find
that the cutoff energy exceeds 10 TeV for typical parameters. In
section 3, we compare the spectra of the delayed emission for
these three emission models of TeV photons and give a more
accurate calculation of the delayed time caused by the angular
spreading by considering recent observations of the extragalactic
infrared background and the theoretic high-redshift infrared
background. Then, in section 4, we study the particular case of
GRB940217, from which delayed emission had already been detected
by EGRET. Finally, we give the conclusions and discussions.

\section{ TeV emission models of GRBs }

\subsection{Cutoff energy of high energy photons from external shocks}
GRBs are thought to be caused by the dissipation, through shocks,
of the kinetic energy of a relativistically expanding fireball
with a Lorentz factor $\Gamma_0\sim 10^2-10^3$. The shocks could
be either {\em internal} (Paczynski \& Xu 1994; Rees \&
M\'{e}sz\'{a}ros 1994) due to collisions between fireball shells
or {\em external} (Rees \& M\'{e}sz\'{a}ros 1992; Dermer,
B\"{o}ttcher \& Chiang 1999) due to the interaction of the
fireball with the external medium. When the relativistic ejecta
encounters the external medium, a relativistic {\em forward} shock
expands into the external medium, and a {\em reverse} shock moves
into and heats the fireball ejecta (Sari \& Piran 1999).  Very
high energy photons can be produced by electron IC and
proton-synchrotron emission in both internal shocks and external
shocks{\footnote{Here the ``external shock" used in this paper
means the shock at the initial phase of the deceleration of the
fireball, not  the later afterglow shock.}}. {There should be a
cutoff in  the high energy gamma-ray spectrum due to the internal
absorption of high energy gamma rays by pair-production in GRBs
(e.g. Baring \& Harding 1997; Totani 1999; Lithwick \& Sari
2001)}. Recently, Guetta \& Granot (2003) argue that a high cutoff
energy for emission from internal shocks needs large value of the
initial Lorentz factor $\Gamma_0$ or variability time $t_v$, but
at the same time, they imply lower values of the peak energy $E_p$
of the synchrotron emission.  So they conclude that TeV photons
from internal shocks can hardly reconcile with typical GRBs, but
may be related with X-ray flashes. { However, this conclusion is
dependent on the assumption that the observed GRB spectral peak is
due to characteristic synchrotron photon energy, but this is not
completely confirmed. For example, Totani (1999) pointed out that
efficient pair-production in GRBs may affect the peak of GRB
spectrum around MeV.} We note that TeV photons could also come
from external shocks and the cutoff energy may be significantly
increased because external shocks have much larger sizes than
those of internal shocks. Below we will first estimate the cutoff
energy for external shocks.

We adopt an analytical approach similar to the one developed for
internal shocks by Lithwick \& Sari (2001) and applied to
afterglows by Zhang \& M\'{e}sz\'{a}ros (2001), but we here apply
to the initial phase of external shocks.  The attenuation optical
depth of high energy photons with softer photons  is (Lithwick \&
Sari 2001)
\begin{equation}
\tau_{\gamma\gamma}=(11/180)\sigma_{\rm T} N({>E_{\rm an}})/4\pi
r_{\rm dec}^2,
\end {equation}
where $N({>E_{\rm an}})$ is the total photon number with energy
above the attenuation threshold energy $E_{\rm an}$, $r_{\rm dec}$
is the deceleration radius at which external shooks take place and
$\sigma_{\rm T }$ is the Thomson cross section. As Zhang \&
M\'{e}sz\'{a}ros (2001), we assume that the emission spectrum
around $E_{\rm an}=h\nu_{\rm an}$ is $L(\nu)=F_\nu 4\pi D^2/(1+z)
\propto \nu^{-\beta}$, then the optical depth is
\begin{equation}
\tau_{\gamma\gamma}=\frac{(11/180) \sigma_{\rm T} F_\nu(\nu_{\rm
an}) d_L^2}{4 \Gamma_0^4 c^2 h \beta t_{\rm dec}},
\end {equation}
where $d_L$ is the source distance, $z$ is the redshift of the
source, $h$ is the Planck constant and $t_{\rm
dec}=10E_{53}^{1/3}n_0^{-1/3}(\Gamma_0/300)^{-8/3}\,{\rm s}$ is
the deceleration timescale of the GRB ejecta. For primary photons
in TeV band , the attenuation threshold energy is $E_{\rm
an}=h\nu_{\rm an}=20(\frac{\Gamma_0}{300})^2(\frac{{\cal
E_\gamma}}{1{\rm TeV}})^{-1}{\rm keV} $. At $h\nu\sim 10{\rm
keV}$, the emission is dominated by the electron synchrotron
radiation from the  external forward shocks (see Fig. 2 in Wang,
Dai \& Lu 2000). The two charactristic frequencies and the peak
flux of the synchrotron spectrum of the external forward shocks
are given by
\begin{equation}
\nu_m^{fs}=4\times10^{20}\left(\frac{p-2}{p-1}\right)^2
\left(\frac{\epsilon_e}{0.5}\right)^2
\epsilon_{B,-2}^{1/2}\left(\frac{\Gamma_0}{300}\right)^4 n_0^{1/2}
{\rm Hz},
\end{equation}
\begin{equation}
\nu_c^{fs}=\frac{10^{17}}{(Y+1)^2}E_{53}^{-1/2}\epsilon_{B,-2}^{-3/2}n_0^{-1}\left(\frac{t_{\rm
dec}}{10 \,\rm s}\right)^{-1/2}{\rm Hz},
\end{equation}
and
\begin{equation}
F_{\nu_m}^{fs}=26D_{L,28}^{-2}\epsilon_{B,-2}^{1/2}E_{53}n_0^{1/2}\,{\rm
mJy},
\end{equation}
respectively, where $\epsilon_e$ and $\epsilon_B$ are the
fractions of the shock energy carried by electrons and magnetic
field respectively, $Y\simeq \sqrt{\frac{\epsilon_e}{\epsilon_B}}$
is the Compton factor and $n$ is the number density of the
external medium. We use the usual notation $a=10^n a_n$ throughout
the paper. Generally we have $\nu_c^{fs}<\nu_{\rm
an}<\nu_{m}^{fs}$, so
\begin{equation}
F_\nu(\nu_{\rm an})=(\frac{\nu_{\rm
an}}{\nu_c^{fs}})^{-1/2}F^{fs}_{\nu_m}=3.8(Y+1)^{-1}\left(\frac{\Gamma_0}{300}\right)^{-1}E_{53}^{3/4}
\epsilon_{B,-2}^{-1/4}n_0^{1/2}t_{\rm dec,
1}^{-1/4}D_{L,28}^{-2}(\frac{\cal E_\gamma}{1\rm TeV})^{1/2}{\rm
mJy}.
\end{equation}
From Eq.(1), finally we  get the cutoff energy where
$\tau_{\gamma\gamma}=1$
\begin{equation}
E_{cut}\sim40\left(\frac{\Gamma_0}{300}\right)^{10/3}E_{53}^{-2/3}\epsilon_{B,-2}^{1/2}
n_0^{-5/6} \,{\rm TeV}.
\end{equation}

\subsection {Spectrum of the primary TeV photons for different emission
models}

One mechanism for the TeV photon production from GRBs is the
electron IC emission in GRB shocks. As the electrons in internal
shocks and external forward shocks are in the fast-cooling regime
and $h\nu_{\rm KN}^{\rm IC}<{\rm TeV}<h\nu_M^{\rm IC}$,   the
energy spectrum of the electron IC emission at TeV band can be
commonly described as
 $ \nu F_{\nu} \propto  \nu^{-p+1/2}$(Guetta \& Granot 2003),
where
 $h\nu_{\rm KN}^{\rm
IC}=\Gamma_0^2 m_e^2 c^4/h\nu_m=22{\rm
GeV}\frac{\Gamma_0}{300}(\frac{h\nu_m}{1{\rm MeV}})^{-1}$ ($\nu_m$
represent the peak frequency of the synchrotron spectrum for
internal or forward shocks) and $h\nu_M^{\rm IC}=\Gamma_0 \gamma_M
m_e c^2$ ($\gamma_M$ is the maximum Lorentz factor of the
electrons accelerated by shocks). On the other hand, for reverse
shocks,  $h\nu_c^{\rm IC} <{\rm TeV} <h\nu_{\rm KN}^{\rm IC}$
(where $\nu_c^{\rm IC}\simeq 2\gamma_c^2 \nu_c$, $\nu_c$ is the
cooling break frequency of the reverse shocks and $\gamma_c$ is
the Lorentz factor of the corresponding electrons), so TeV
spectrum is given by $\nu F_{\nu} \propto \nu^{-p/2+1}$ (Sari \&
Esin 2001).

In the region where the electrons are accelerated, protons may be
also accelerated up to ultra-high energies $>10^{20}{\rm eV}$
(Waxman 1995; Vietri 1995), producing a spectrum characteristic of
Fermi mechanism $dN_p/dE_{p}\propto E_{p}^{-p}$. The possibility
of energetic protons, accelerated in both internal shocks and
external shocks, producing $\sim {\rm TeV}$ gamma rays by
synchrotron emission has been discussed by a number of authors
(Vietri 1997; B\"{o}ttcher \& Dermer 1998; Totani 1998a,b). For
external shocks, the post-shock magnetic field is
$B=(32\pi\epsilon_B \Gamma_0^2 nm_p c^2)^{1/2}$, and the energy of
the synchrotron photons is given by
\begin{equation}
E_{p-syn}=\frac{\Gamma_0\Gamma_p^2 e hB}{2\pi m_p c}=3{\rm TeV}
E_{p,21}^2\epsilon_B^{1/2}n_0^{1/2},
\end{equation}
where $\Gamma_p$ and $E_p$ are respectively the Lorentz factor and
energy of the protons. Totani (1998a,b) argues that the protons
can be accelerated up to $10^{20}-10^{21}{\rm eV}$ for
$\Gamma_0=100-1000$, so we expect $E_{p-syn}$ can extend to TeV
band for $\epsilon_B n\sim 1$. The energy spectrum from
proton-synchrotron radiation is $\nu F_\nu\propto \nu^{(3-p)/2}$,
which is distinct from the electron IC emission spectrum we
discussed above.

\section{Spectrum, duration and intensity of the delayed emission}

\subsection{Spectrum of the delayed emission}
TeV gamma-rays emitted from extragalactic sources may collide with
diffuse cosmic infrared background (CIB) photons, leading to
secondary $e^+e^-$ pairs. { The pair production optical depth
$\tau_{\gamma\gamma}$ depends on the spectral energy distribution
and the intensity of the CIB, which is currently not well-known.
Because of the high redshift of cosmological GRB sources,
$\tau_{\gamma\gamma}$ also depends on the evolution of the CIB
with the redshift, which  also remains uncertain. Despite of these
uncertainties, calculations based on the theoretic models ( e.g.
de Jager \& Stecker 2002; Aharonian et al. 2002 ) of CIB and
modelling of observations of the TeV blazar H1426+428 (Aharonian
et al. 2002; Costamante et al. 2003) show that
$\tau_{\gamma\gamma}$ is significantly larger than unity for
photons with ${\cal E_\gamma}\ga500{\rm GeV}$ from extragalactic
GRBs with redshift $z\ga0.3$ (also see Totani et al. 2002 in which
a somewhat lower optical depth is obtained). On the other hand,
${\cal E_\gamma}$ from the synchrotron radiation of protons or
electron IC emission may extend to a few TeV. So we here choose to
study the primary very high-energy photons with energy in the
range of 0.5-5 TeV that are almost totally absorbed by the CIB
photons.} We assume a general form for the energy spectrum of the
primary high energy emission at TeV band: $\nu L_\nu\propto
\nu^{-\alpha}$. The photon spectrum is accordingly $N_\nu\propto
\nu^{-(\alpha+2)}$ and the spectrum of the secondary pairs is then
$dN_e/d\gamma_e\propto \gamma_e^{-(\alpha+2)}$. The secondary
pairs would boost the CMB photons to higher energy by IC
scattering. The scattered photons (or delayed photons) will have a
characteristic energy
\begin{equation}
\varepsilon=\frac{4}{3}\gamma_e^2 <\epsilon>=0.8 (\frac{{\cal
E}_\gamma}{1{\rm Tev}})^2 {\rm GeV},
\end{equation}
where $<\epsilon>=2.7kT_{\rm CMB}$ is the mean energy of the CMB
photons and $\gamma_e$ is the Lorentz factor of the secondary
pairs resulted from a primary photon with energy ${\cal
E_\gamma}$. { So, for ${\cal E_\gamma}$ in the range of 0.5-5 TeV,
the energies of the scattered photons are in the range
200MeV-20GeV correspondingly.}

The time integrated spectrum of the scattered CMB photons should
be
\begin{equation}
\frac{dN_\varepsilon}{d\varepsilon}\propto
\varepsilon^{-(\alpha+4)/2}
\end{equation}
(Blumenthal \& Gould 1970, Dai \& Lu 2002). { Strictly speaking,
this form holds only when all the TeV photons are absorbed
locally, i.e. their production mean free path $R_{\rm pair}$
should be much smaller than the luminosity distance of the
sources; otherwise, TeV photons of different energy may be
absorbed at different redshift $z_{\rm pair}$, causing  the
observed delayed photons energy shifted from Eq.(9) by a factor
$1-(\frac{1+z_{\rm pair}}{1+z})^2$.  But, for the calculated
values of $R_{\rm pair}$ in the next subsection, we find that this
factor is within 15{\%} and so this form holds with a good
approximation. It is important to note that if we choose the part
of TeV photons that are totally absorbed locally, the spectrum of
the delayed emission is independent of the poorly-known CIB.} For
three different TeV models, the spectra of the scattered (delayed)
emission are different, as presented in Table 1. From Table 1, we
can see that the spectra of the delayed emission are significantly
different from each other and we can therefore use this difference
to constrain the emission mechanism of the primary TeV photons.

\subsection{The duration  of the
delayed emission }

The duration of the scattered CMB photons should be the maximum of
three timescales: $\tau_1$--the observed IC cooling life time of
the secondary electrons, $\tau_2$--the time scale caused by the
deflection of the electrons due to the IGMF and $\tau_3$--the
angular spreading time (Dai \& Lu 2002; Dai et al. 2002). Dai \&
Lu (2002) have derived
\begin{equation}
\tau_1=\frac{3m_e c}{8 \gamma_e^3 \sigma_T u_{\rm
CMB}}=37\left(\frac{{\cal E_\gamma}}{\rm 1 TeV}\right)^{-3}\,{\rm
s}=37\left(\frac{\varepsilon}{\rm 0.8 GeV}\right)^{-3/2}\,{\rm s},
\end{equation}
where $u_{\rm CMB}$ is the energy density of the CMB photons and
$\gamma_e$ is the Lorentz factor of the secondary electrons and it
relates to the energy of the primary TeV photons by $\gamma_e=10^6
({{\cal E_\gamma}/1{\rm TeV}})$. The time scale caused by the
deflection of the electrons due to the IGMF is given by
\begin{equation}
\tau_2=6.1\times10^3 \left(\frac{{\cal E_\gamma}}{\rm 1
TeV}\right)^{-5}\left(\frac{B_{\rm IGMF}}{10^{-20}\rm G}\right)^2
\,{\rm s}=6.1\times10^3 \left(\frac{\varepsilon}{\rm 0.8
GeV}\right)^{-5/2}\left(\frac{B_{\rm IGMF}}{10^{-20}\rm
G}\right)^2 \,{\rm s}
\end{equation}

To know $\tau_3$, we must know the mean free path $R_{\rm pair}$
of the very high energy photons in the extragalactic IR
background, which depends on the intensity of the IR background.
Electron-positron pair creation due to the interaction of a
$\gamma$-ray photon of energy ${\cal E_\gamma}$ with a softer
photon of energy $\epsilon$ can take place provided that ${\cal
E_\gamma} \epsilon (1-{\rm cos}\theta)\ge2(m_e c^2)^2$, where
$\theta$ is the encounter angle of the two photons. For a fixed
$\gamma$-ray energy ${\cal E_\gamma}$, the pair production cross
section $\sigma$ rises steeply from the threshold $\epsilon_{\rm
th}$, has a maximum value equal to $0.26\sigma_{\rm T}$ at
$\epsilon=2m_e^2 c^4/{\cal E_\gamma}$ and then falls off as
$\epsilon^{-1}$ for $\epsilon>\epsilon_{\rm th}$. Because of the
peaked cross section, collisions will preferentially take place
between $\gamma$-ray photons of energy ${\cal E_\gamma}$ and soft
photons with energy $\sim 2m_e^2 c^4/{\cal E_\gamma}$. So, for
${\cal E_\gamma}$ in the range of 0.5--5 TeV, the wavelengths of
the softer photons with which pair production preferentially takes
place are in the range $1.2-12\mu m$.

Since GRBs are at cosmological distances and CIB evolves with the
redshift (Salamon \& Stecker 1998), $R_{\rm pair}$ should depend
on the redshift of the GRBs. For simplicity, we will discuss three
representative cases with a) { $z=0.3$}, b)$z=1$ and c) $z=3$.

{\em Case a ({ $z=0.3$}):} { It can be assumed that CIB is in
place by a time corresponding to $z=0.3$, so the CIB photon number
density at $z=0.3$ is $(1+z)^3$ higher than the value at $z=0$.
Observations of CIB show that its spectral energy distribution has
a peak around $1-2\mu m$ and a valley at mid-IR band.  In the
wavelength range of $\sim1.2-12\mu m$ that we are interested in,
the number densities of the CIB photons can be approximately
shaped by $\bar n(\epsilon)\epsilon\propto \epsilon^{-k_1}$ with
$-k_1\simeq0$ (Coppi \& Aharonian 1999; { Aharonian et al. 2002}),
which results in a nearly constant $R_{\rm pair}$. }The observed
CIB flux $J_\nu$ at $2.2\mu m$ is of the order of $\sim 10 \,{\rm
nW/m^2/sr}$ (Wright \& Johnson 2001; Wright 2003), { so the number
density of the CIB photons  at $z=0$ }is
\begin{equation}
\bar n(\epsilon)\epsilon|_{\lambda=2.2\mu m} =\frac{4\pi J_\nu}{c
\epsilon} =0.45\times10^{-2} {\rm cm^{-3}}.
\end{equation}
Then we obtain {  the mean free path for TeV photons at $z=0.3$}
\begin{equation}
R_{\rm pair}=\frac{1}{0.26\sigma_{\rm T}\bar
n(\epsilon)\epsilon(1+z)^3}=0.55\times 10^{27} (\frac{{\cal
E_\gamma}}{1\rm TeV})^{-k_1} \, {\rm cm}
\end{equation}
and
\begin{equation}
\tau_3 (z=0.3)=\frac{R_{\rm pair}}{2\gamma_e^2 c}=1.0\times 10^4
(\frac{{\cal E_\gamma}}{1\rm TeV})^{-2-k_1}\, {\rm s}=1.0\times
10^4 (\frac{\varepsilon}{0.8\rm GeV})^{-1-k_1/2}\, {\rm s}
\end{equation}

{\em Case b ($z=1$): }  Salamon \& Stecker (1998) have derived the
intergalactic comoving radiation energy density as a function of
wavelength in the range $10^{-2}$--$2.5\mu m$ for several fixed
redshifts by considering the stellar emissivity with and without
metallicity. We shall assume that the power-law form
$n(\epsilon)\epsilon\propto \epsilon^{-k_2}$ also holds in the
range 1.2-12$\mu m$ for $z=1$, but the intensity of CIB is lower
than that at $z\simeq0$, according to the calculation result of
Salamon \& Stecker (1998). At $\lambda=2.2\mu m$, { the
intergalactic comoving radiation energy density $U_\nu$ at $z=1$}
is about $9\times10^{-30} {\rm erg Hz^{-1} cm^{-3}}$, so
\begin{equation}
n(\epsilon, z=1)\epsilon=\frac{U_\nu}{h}=1.4\times 10^{-3}
(\frac{\lambda}{2.2\mu m})^{k_2}.
\end{equation}
Finally, we obtain $R_{\rm pair}$ and the delay time $\tau_3$ for
bursts at $z=1$:
\begin{equation}
R_{\rm pair}=\frac{1}{0.26\sigma_{\rm T}n(\epsilon, z=1
)\epsilon}=4\times 10^{27} (\frac{{\cal E_\gamma}}{1\rm
TeV})^{-k_2} \, {\rm cm}
\end{equation}
\begin{equation}
\tau_3(z=1)=6.67\times 10^4 (\frac{{\cal E_\gamma}}{1\rm
TeV})^{-2-k_2}\, {\rm s}=6.67\times 10^4
(\frac{\varepsilon}{0.8\rm GeV})^{-1-k_2/2}\, {\rm s}
\end{equation}

{\em Case c ($z=3$):} The CIB radiation energy density is even
lower. At $\lambda=2.2\mu m$, $U_\nu$ is about
$3\times10^{-30}{\rm erg Hz^{-1} cm^{-3}}$. So
\begin{equation}
\tau_3(z=3)=2.2\times 10^5 (\frac{{\cal E_\gamma}}{1\rm
TeV})^{-2-k_3}\, {\rm s}=2.2\times 10^5 (\frac{\varepsilon}{0.8\rm
GeV})^{-1-k_3/2}\, {\rm s}
\end{equation}
where $n(\epsilon)\epsilon\propto \epsilon^{-k_3}$
($\lambda=1.2-12\mu m$) for CIB at $z=3$. If we also consider the
time dilation due to redshift, the delay time is even longer by a
factor of $1+z$.

To know the values of $k_2$ and $k_3$ in the range
$\lambda=1.2-12\mu m$, we need to know the intergalactic comoving
radiation energy density beyond $\lambda= 2.5\mu m$ (Salamon \&
Stecker 1998). But we expect that the absolute value of $k_2$ or
$k_3$ is small, since they reflect the characteristic shape of the
starlight spectrum which in the wavelength band between 1 and
several microns behaves as $\nu I(\nu)\propto\lambda ^\beta $ with
$\beta\sim-1$.

Comparing these three timescales, we know that the observed IC
life time $\tau_1$ is always much smaller than the other two
timescales and therefore not related to the delay time. $\tau_2$
may be comparable to $\tau_3$ at $\varepsilon=0.8{\rm GeV}$ if
$B_{\rm IGMF}\sim 3\times 10^{-20}{\rm G}$ for bursts at $z\simeq
1$. By now, very little is known about the IGMF. To interpret the
observed $\sim\mu {\rm G}$ magnetic fields in galaxies and X-ray
clusters, the seed fields required in dynamo theories could be as
low as $10^{-20}{\rm G}$ (Kulsrud et al. 1997; Kulsrud 1999).
Theoretical calculation of primordial magnetic fields show that
these fields could be of order $10^{-20}{\rm G}$ or even as low as
$10^{-29}{\rm G}$, generated during the cosmological QCD or
electroweak phase transition, respectively (Sigl, Olinto \&
Jedamzik 1997). So, if we know which timescale is responsible for
the delay time, we can constrain the strength of the IGMF.

A way to distinguish which mechanism causes the time delay $\Delta
t $ is to examine the dependence of the maximum time delay $\Delta
t (\varepsilon)$ of the scattered photons  on their energy
$\varepsilon$. From the expression of $\tau_2$ (Eq.12) or $\tau_3$
(Eqs. 15, 18 or 19), we can obtain the dependence of the maximum
time delay on photon energy. If the delay time is dominated by
$\tau_2$, $\Delta t(\varepsilon)\propto \varepsilon^{-5/2}$, while
$\Delta t(\varepsilon)\propto \varepsilon^{-(1+\kappa)}$ if it is
dominated by $\tau_3$, where $\kappa=0, k_1/2, k_2/2$ for bursts
at $z=0.3, 1, 3 $ respectively.

Regardless which of the two timescales is responsible for the time
delay, from the expressions of $\tau_2$ and $\tau_3$, we know that
softer photons tend to have larger amounts of delay. Another
feature characteristic of our model for the delayed emission is
that the flux of the delayed emission is roughly a constant over
the whole duration. These may be used to tell our model from other
models for the delayed emission from GRBs. For example, Zhang \&
M\'{e}sz\'{a}ros (2002) proposed that the electron IC emission
from afterglow shocks could produce a delayed GeV component. But
in their model, the GeV flux rises first to a peak and then
declines in a power law manner as $F_\nu\propto t^{(11-9p)/8}$
(Zhang \& M\'{e}sz\'{a}ros 2002).

\subsection{Intensity of the delayed emission}
We  assume that the burst energy in the TeV energy range from
${\cal E}_{\gamma,1}=h\nu_{1}=0.5 {\rm TeV}$ to ${\cal
E}_{\gamma,2}=h\nu_{2}=5 {\rm TeV}$ is a fraction $f$ of the burst
energy in the BATSE energy range $E_{\rm B}$, i.e.
$\int_{\nu_1}^{\nu_2} L_\nu d\nu \Delta t=fE_{\rm B}$ , where
$L_\nu$ is the luminosity per frequency and $\Delta t$ is the
duration of the primary TeV emission, which is considered to be
comparable to the prompt GRB duration in the BATSE band.

The energy of the primary photons from $h\nu_{1}$ to $h\nu_{2}$ is
essentially redistributed to the scattered photons from
$\varepsilon_1=200 {\rm MeV}$ to $\varepsilon_2=20 {\rm GeV}$,
i.e.
\begin{equation}
\int_{\varepsilon_1}^{\varepsilon_2}\varepsilon\frac{dN_\varepsilon}{d\varepsilon}
=\int_{\nu_1}^{\nu_2} L_\nu d\nu \Delta t=fE_{\rm B}.
\end{equation}
So, the observed, time integrated (over the total duration)
differential energy spectrum is

\begin{equation}
\begin{array}{ll}
\varepsilon^2 \frac{dN_\varepsilon}{d\varepsilon}\Delta t
=\frac{\alpha}{2[1-(\nu_2/\nu_1)^{-\alpha}]}\frac{f E_{\rm
B}}{4\pi d_L^2}(\frac{\varepsilon}{\varepsilon_1})^{-\alpha/2}
=f_1f \frac{f E_{\rm B}}{4\pi
d_L^2}(\frac{\varepsilon}{\varepsilon_1})^{-\alpha/2}\\
=6\times10^{-7}\cdot
11^{-\frac{\alpha}{2}}(\frac{f_1}{0.25})f_{-2}E_{\rm
B,53}d_{L,28}^{-2}(\frac{\varepsilon}{0.8 \rm
GeV})^{-\alpha/2}{\rm erg cm^{-2}} ,
\end{array}
\end{equation}
where $f_1=\frac{\alpha}{2[1-(\nu_2/\nu_1)^{-\alpha}]}$ are
respectively 0.13, 0.25 and 0.85 for three kinds of TeV spectrum
for $p=2.2$. The fraction $f$ of the burst energy in TeV band  is
an unknown factor and may depend on the emission model and many
unknown parameters like $\epsilon_e$ and $\epsilon_B$. For the
model of the IC emission from the reverse shocks, we estimate
$f\sim0.01$ for typical shock parameters (see Fig. 2 in Wang, Dai
\& Lu 2001), which is consistent with the observational result
that the energy in the observed delayed emission from GRB940217 is
about a factor of 0.01 of the burst energy in the BATSE range.

The EGRET detector has fluence threshold of $\sim
2.1\times10^{-6}{\rm erg cm^{-2}}$ for a short integration time
regime ($t<1.7\times10^3{\rm s}$) and of $\sim
2.1\times10^{-6}{\rm erg cm^{-2}}(t/1.7\times10^3 {\rm
s})^{\frac{1}{2}}$ for long-term observations. Therefore, EGRET
could detect delayed GeV emission only from strong GRBs, such as
GRB940217. However, the future detector GLAST is much more
sensitive.  The fluence threshold for
 GLAST is roughly $\sim 4\times10^{-7} ({t}/{10^5 \rm
s})^{1/2} {\rm erg cm^{-2}}$ for a long integration time regime
(exposure time $t\ga 10^5 {\rm s}$) and $\sim 4\times 10^{-7}{\rm
erg cm^{-2}}$ for a short integration time (Gehrels \& Michelson
1999; Zhang \& M\'{e}sz\'{a}ros 2001). So, we expect that GLAST
could detect delayed emission from  typical GRBs with $E_{\rm
B}=10^{53}{\rm erg}$ at $d_L=10^{28}{\rm cm}$.

\section{GRB940217}

GRB940217 is a very strong burst with a total fluence above 20 keV
of $(6.6\pm2.7)\times10^{-4}{\rm erg cm^{-2}}$ and a duration of
$\sim 180$  s in the BATSE range (Hurley et al. 1994). It has the
third-largest fluence of $\sim 800$ BATSE bursts up to the
detection time of this burst. During the period of the low energy
emission, i.e. the first $\sim 180$ s, EGRET detected 10 photons
with energies ranging from a few tens of MeV to a few GeV and the
flucence in this range is $\sim 2\times10^{-5}{\rm erg cm^{-2}}$.
Most strikingly, an additional 18 high-energy photons were
recorded for $\sim 5400 $  s following this, including  an 18-GeV
photon and { other 36-137 MeV photons}. The fluence of the delayed
emission was measured to be $7\times 10^{-6}{\rm erg cm^{-2}}$ in
the energy range 30MeV-3GeV. { Among many models for  for this
delayed high-energy emission from GRBs (e.g. M\'{e}sz\'{a}ros \&
Rees 1994; Katz 1994; Plaga 1995;  Totani 1998a,b; Wang et al.
2002), one is that the delayed emission is the result of the
electromagnetic cascade of the TeV photons and the inverse Compton
scattering of the CMB radiation (Cheng \& Cheng 1996).

{ Although this model suggests that soft photons tend to have
larger amounts of delay time, very long observation time is needed
to detect this effect for delayed photons with energy $\la $ 100
MeV, as shown by expressions of $\tau_3$. For GRB940217, the lack
of high energy photons within 36-137 MeV should occur at around
$10^5-10^6 {\rm s}$ after the keV burst, while the observation of
EGRET takes only $\sim 5000\, {\rm s}$. Thus, no correlation
between time delay and photon energy could be found in the EGRET
observations.  }

In this model, the 36-137 MeV photons should come from the cascade
process of photons with energies  ${\cal E_\gamma}< 0.5 {\rm TeV}
$. Since this part of high energy photons may not be totally
absorbed by IR photons, especially if this burst is at a low
redshift, we get $f\ga0.01$ for this burst.} Up to the detection
time of GRB940217, EGRET has had some exposure to about 150 BATSE
bursts (Hurley et al. 1994). Since GRB940217 was the
third-strongest one of $\sim 800$ BATSE burst and had fluence of
$(6.6\pm2.7)\times10^{-4}{\rm erg cm^{-2}}$, we estimate that only
$\sim 1$ burst with fluence $F\ga 2\times10^{-4}{\rm erg cm^{-2}}$
had been exposed by EGRET. If $f\sim 0.01$, then only $\sim 1$
burst will have delayed emission sufficiently strong to be
detected by EGRET, which is consistent with the only one detection
of the delayed emission from GRBs by EGRET.

{ In the energy range $\epsilon=200{\rm MeV}-20{\rm GeV}$, there
is only one delayed photon detected  from GRB940217 and therefore
we have no reliable photon spectrum for the delayed emission of
this energy range. Below this energy, the delayed photon spectrum
may deviate from the form Eq.(10), as the original very high
energy photons corresponding to this part of delayed emission may
be partially absorbed by the CIB. Moreover, the expected theoretic
spectrum Eq. (10) is a time integrated spectrum over the whole
duration of the delayed emission, while GRB940217 was observed in
a limited time. So, we could not constrain the origin of its TeV
photons by comparing the form Eq.(10) with the the photon spectrum
of the delayed emission from GRB940217, although a numerical
approach taking account of the gamma-ray absorption effect might
be feasible. However, we expect that this method has a promising
prospect in the future GLAST era because of the significantly
increased sensitivity of this detector, with the delayed emission
detected from much more GRBs and the spectrum of photons
$\epsilon\ga 200{\rm MeV}$ well-determined.}

\section{Summary}
In previous papers (Cheng \& Cheng 1996; Dai \& Lu 2002), we have
suggested that the very high energy photons from cosmological GRBs
may collide with cosmic IR background photons, leading to
electron/positron pair production. Inverse Compton scattering of
the pairs off CMB photons will produce delayed MeV-GeV emission.
In this paper, we extended our previous works  by the following
points:

1)We suggest that TeV photons could also come from GRB external
shocks. Compared with internal shocks, TeV photons from external
shocks suffer little attenuation due to pair production with
softer photons in the bursts.

2)There are a few emission models suggested for the TeV photons
from GRBs, such as the proton-synchrotron radiation, the electron
IC emission from  external reverse shocks, and electron IC
emission from intern shocks or external forward shocks. { In this
paper, we suggest that the spectrum of the delayed emission
resulted from the TeV photons that was totally absorbed locally by
the intergalactic IR background radiation  could help to
constraint the emission model of TeV photons from GRBs. Because
this part of TeV photons are absorbed locally, the spectrum of the
delayed emission is independent of the poorly-known CIB.} { Since
our treatment here is mainly analytic, a further numerical study
to confirm this idea might be useful and necessary. }

3)The time delay could be caused by the angular spreading effect
of the scattered microwave photons or deflection of the secondly
pairs due to IGMF.  We present a more accurate calculation of the
delay time caused by the angular spreading effect of the secondary
electrons by considering recent observations of the extragalactic
IR background and the theoretic prediction of the high-redshift IR
background. By examining the dependence of the delay time on the
photon energy, i.e. $\Delta t (\varepsilon)\propto
\varepsilon^{-\delta}$, one can tell which of the two timescales
is responsible for the delay time. For the delay time caused by
the angular spreading effect, $\delta\simeq 1$, while $\delta=2.5$
otherwise.

Finally, we would like to point out that this model predicts a
roughly constant flux for the delayed emission over the whole
delay time and  that soft photons tend to have larger amounts of
delay, which constitute distinguished features to differentiate
our model from other models.

 {\acknowledgments { We are grateful to the referee for his/her valuable comments
 and suggestions which improved the paper.}
This work was supported by  the Special Funds for Major State
Basic Research Projects, the National 973 Project, the National
Natural Science Foundation of China under grants 10233010 and
10221001, the Nanjing University Talent Development Foundation
and a RGC grant of Hong Kong government.}

\clearpage

\begin{center}
\begin{table*}[ht!]
\begin{center}
\begin{tabular}{|c|c|c|}
\hline & $-(\alpha+2)$ (values for $p=2.2$) & $-\beta$ (values for $p=2.2$) \\

\hline\hline proton-synchrotron &
$-\frac{p+1}{2}$ \,\,$(-1.6)$  & $-\frac{p+5}{4}$ \,\,$(-1.8)$\\
\hline
IC from reverse shocks& $-\frac{p+2}{2}$ \,\,$(-2.1)$ & $-\frac{p+6}{4}$ \,\,$(-2.05)$ \\
 \hline
IC from forward (or internal) shocks & $-(p+\frac{3}{2})$ \,\,$(-3.7)$ & $-\frac{2p+7}{4}$\,\, $(-2.85)$  \\
 \hline

\end{tabular}
\end{center}
\par
\label{} \caption{Comparison of the spectra of the primary TeV
photons and the delayed photons for different emission models of
TeV photons. $-(\alpha+2)$ ($\frac{dN_{\cal E_\gamma}}{d\cal
E_\gamma}\propto {\cal E_\gamma}^{-(\alpha+2)}$) is the spectral
index of the photon spectrum of the primary TeV photons and
$-\beta$ ($\frac{dN_\varepsilon}{d\varepsilon}\propto
\varepsilon^{-\beta}$) is the spectral index of the photon
spectrum of the delayed MeV-GeV photons. }
\end{table*}
\end{center}


\begin{thebibliography}{99}
\bibitem[]{697}
{ Aharonian, F., et al. 2002, A\&A, 384, L23}
\bibitem[]{669}
{ Amenomori, M., et al. 1996, A\&A, 311, 919}
\bibitem[]{701}
Atkins, R., et al., 2000, ApJ, 533, L119
\bibitem[]{777}
{ Baring, M.G. \& Harding, A. K., 1997, ApJ, 491, 663}
\bibitem[]{671}
Blumenthal, G. R. \& Gould, R. J., 1970, Rev. Mod. Phys., 42, 237
\bibitem[]{673}
B\"{o}ttcher, M. \& Dermer, C. D., 1998, ApJ, 499, L131
\bibitem[]{675}
Cheng, L. X. \& Cheng, K. S., 1996, ApJ, 459, L79
\bibitem[]{677}
Coppi, P. S. \& Aharonian, F. A., 1999, Astropar. Phys., 11, 35
\bibitem[]{711_a}
{ Costamante, L., Aharonian, F., Ghisellini, G., Horns, D. 2003,
New Astronomy Reviews, 47, 677 }
\bibitem[]{679}
Dai, Z. G. \& Lu, T., 2001, ApJ, 551, 249
\bibitem[]{681}
Dai, Z. G. \& Lu, T., 2002, ApJ, 580, 1013
\bibitem[]{683}
Dai, Z. G., Zhang, B., Gou, L. J., M\'{e}sz\'{a}ros, P. and
Waxman, E., 2002, ApJ, 580, L7
\bibitem[]{720}
{ De Jager, O. C. \& Stecker, F. W., 2002, ApJ, 566, 738}
\bibitem[]{685}
Dermer, C. D., B\"{o}ttcher, M. \& Chiang, J., 1999, ApJ, 515, L49

\bibitem[]{687}
Gehrels, N. \& Michelson, P., 1999, Astropar. Phys., 11, 277
\bibitem[]{689}
Guetta, D., \& Granot, J., 2003, ApJ, 585, 885
\bibitem[]{691}
Hurley, K. et al. 1994, Nature, 371, 652
\bibitem[]{730}
{ Katz, J. I., 1994, ApJ, 432, L27}
\bibitem[]{693}
Kulsrud, R., Cowley, S. C., Gruzinov, A. V., \& Sudan, R. N.,
1997, Phys. Rep., 283, 213
\bibitem[]{696}
Kulsrud, R., 1999, ARA\&A, 37, 37
\bibitem[]{698}
Lithwick, Y. \& Sari, R., 2001, ApJ, 555, 540
\bibitem[]{700}
Madau, P. \& Phinney, E. S., 1996, ApJ, 456, 124
\bibitem[]{741_a}
{ M\'{e}sz\'{a}ros, P. \& Rees, M. J. 1994, MNRAS, 269, L41 }
\bibitem[]{702}
Paczy\'{n}ski, B. \& Xu, G., 1994, ApJ, 427, 708
\bibitem[]{704}
Plaga, R., 1995, Nature, 374, 430
\bibitem[]{747_a}
{ Padilla, L., et al. 1998, A\&A, 337, 43}
\bibitem[]{706}
Poirier, J., et al., 2003, Phys. Rev. D, 67, 042001
\bibitem[]{709}
Rees, M. J. \& M\'{e}sz\'{a}ros, P. 1992, MNRAS, 258, P41
\bibitem[]{711_b}
Rees, M. J. \& M\'{e}sz\'{a}ros, P., 1994, ApJ, 430, L93
\bibitem[]{713}
Salamon, M. H. \& Stecker, F. W., 1998, ApJ, 493, 547
\bibitem[]{715}
Sari, R. \& Piran, T., 1999, ApJ, 520, 641
\bibitem[]{717}
Sari, R. \& Esin, A. A. 2001, ApJ, 548, 787
\bibitem[]{719}
Sigl, G., Olinto, A. V. \& Jedamzik, K., 1997, Phys. Rev. D, 55,
4582
\bibitem[]{722}
Sommer, M., et al., 1994, ApJ, 422, L63
\bibitem[]{724}
Stecker, F. W., De Jager, O. C. \& Salamon, F. W., 1992, ApJ, 390,
L49
\bibitem[]{727}
Totani, T., 1998a, ApJ, 502, L13
\bibitem[]{729}
Totani, T., 1998b, ApJ, 509, L81
\bibitem[]{775}
{ Totani, T., 1999, MNRAS, 307, L41}
\bibitem[]{773}
{ Totani, T., Takeuchi, T., 2002, ApJ, 570, 470}
\bibitem[]{731}
Vietri, M., 1995, ApJ, 453, 883
\bibitem[]{733}
Vietri, M., 1997, Phys. Rev. Lett., 78, 4328
\bibitem[]{735}
Wang, X. Y., Dai, Z. G. \& Lu, T., 2001, ApJ, 556, 1010
\bibitem {}
{ Wang, X. Y., Dai, Z. G. \& Lu, T., 2002, MNRAS, 336, 803}
\bibitem[]{737}
Waxman, E., 1995, Phys. Rev. Lett., 75, 386
\bibitem[]{739}
waxman, E. \& Bahcall, J., 1997, Phys. Rev. Lett., 78, 2292
\bibitem[]{741_b}
Waxman, E. \& Bahcall, J., 2000, ApJ, 541, 707
\bibitem[]{743}
Wright, E. \& Johnson B., 2001,  astro-ph/0107205
\bibitem[]{745}
Wright, E. 2003, astro-ph/0306058
\bibitem[]{747_b}
Zhang, B., \& M\'{e}sz\'{a}ros, P., 2001, ApJ, 559, 110

\end{thebibliography}
\end{document}